\documentclass[12pt]{article}

\usepackage{scicite}

\usepackage{times}

\usepackage{graphicx}
\usepackage{amsmath}

\newenvironment{sciabstract}{%
\begin{quote} \bf}
{\end{quote}}

\title{Propagating bands of plastic deformation in a metal alloy as critical avalanches}

\author{
Tero Mäkinen,$^{1\ast}$ Pasi Karppinen,$^2$ Markus Ovaska,$^1$\\ Lasse Laurson,$^3$ Mikko J. Alava$^{1,4}$\\
\\
\normalsize{$^1$Department of Applied Physics, Aalto University,}\\
\normalsize{P.O. Box 11100, FI-00076 Aalto, Espoo, Finland}\\
\normalsize{$^2$ProtoRhino Ltd, Betonimiehenkuja 5C, FI-02150 Espoo, Finland}\\
\normalsize{$^3$Computational Physics Laboratory, Tampere University,}\\
\normalsize{P.O. Box 692, FI-33014 Tampere, Finland}\\
\normalsize{$^4$NOMATEN Centre of Excellence, National Centre of Nuclear Research,}\\
\normalsize{A. Soltana 7, 05-400 Otwock--\'Swierk, Poland}\\
\\
\normalsize{$^\ast$To whom correspondence should be addressed; E-mail:  tero.j.makinen@aalto.fi.}
}

\date{}

\begin{document} 


\baselineskip24pt


\maketitle


\begin{sciabstract}
 The plastic deformation of metal alloys localizes in the Portevin--Le Chatelier effect in bands of different types, including propagating, or type ‘A’ bands, usually characterised by their width and a typical propagation velocity. This plastic instability arises from collective dynamics of dislocations interacting with mobile solute atoms but the resulting  sensitivity to the strain-rate lacks fundamental understanding. Here we show, by employing high-resolution imaging in tensile deformation experiments of an aluminum alloy, that the band velocities exhibit large fluctuations. Each band produces a velocity signal reminiscent of crackling noise bursts observed in numerous driven avalanching systems from propagating cracks in fracture to the Barkhausen effect in ferromagnets. The statistical features of these velocity bursts including their average shapes and size distributions obey predictions of a simple mean-field model of critical avalanche dynamics. Our results thus reveal a novel paradigm of criticality in the localization of deformation.
\end{sciabstract}


\section*{Introduction}

Complexity in materials deformation is important for engineering and involves fundamental non-equilibrium physics. Such phenomena are encountered when samples are loaded beyond the regime of linear, elastic response. Then,
metals yield and the plastic deformation prior to failure is now known to exhibit very complex properties on various 
scales in time and space \cite{Zaiser2006,Papanikolaou2018,crackling_alava_2014,vu2019compressive}. 
The challenges this brings up range from avalanches of plastic deformation to the statistical fluctuations of 
the yield stress in finite samples to deformation localization.
A typical manifestation of localization is the appearance of shear bands and here we study the Portevin--Le Chatelier effect \cite{old_lechatelier_1909,old_portevin_1923}.\\

The PLC effect implies the creation of deformation bands in a sample 
(Fig. 1) when it is loaded beyond the yield point: such bands nucleate, 
and may or may not propagate depending on the class of PLC instability present \cite{Hahner2002,Lebyodkin1996}
(in the common classification type A denotes propagating and types B and C nonpropagating bands). 
The deformation bands are accompanied by material instabilities, in the case of tensile tests stress-drops which then produce serrated stress-strain curves (Fig. 1b).
This kind of Strain-Rate Sensitivity \cite{wilcox1966serrated,Rajesh2000} (SRS) arises 
as a strain-rate dependent phenomenon; moreover its character and 
presence are dependent on the temperature. The PLC effect is attributed 
to Dynamic Strain Aging \cite{Cottrell1949,fressengeas2005dynamic,zaiser1997oscillatory} (DSA), and the crucial physics 
is in the interaction of the dislocations as the fundamental carriers of 
plastic deformation with the solute atoms in the alloy \cite{const_mccormick_1972,van1975theory,zhao2020simulating}. On the mesoscopic level, 
theories of increasing complexity have been proposed such that they would 
account for the necessary dislocation physics:  elementary classes of 
immobile and ''aging'', solute bound dislocations, and mobile ones 
producing plastic deformation. Such models and a multitude of experiments 
have been recently introduced to explore the physics of the PLC effect: 
phases in the band nucleation \cite{dic_tong_2005,Halim2007,Casarotto2009,Klusemann2015}
and dynamics including serrations in the stress-strain curves 
\cite{Lebyodkin1996,Rajesh2000,Bharathi2001,Chihab2003,Lebyodkin2005,Chatterjee2008,ae_kumar_2015}, acoustic emission \cite{Chmelik2007,Shashkov2012a,Lebyodkin2013,ae_kumar_2015} 
from the effect and so forth.

\section*{Results}

Here we take a fundamentally different approach of coarse-graining, where the bands are reduced to zero-dimensional "particles". This amounts to
studying the propagation velocity signals $v_b(t)$ of each individual 
propagating ('type A') band during a deformation experiment. Our
high-resolution experiments based on speckle imaging of the deforming
sample (see Fig. 1 and Methods for details) reveal that
the $v_b(t)$ signals are reminiscent of crackling noise bursts found
in numerous driven systems ranging from propagating cracks \cite{laurson2013evolution} and
fluid fronts invading porous media \cite{rost2007fluctuations} to the jerky field-driven
motion of domain walls in ferromagnets \cite{durin2006science,durin2000scaling,zapperi1998dynamics} (see Fig. 1c). 
This is in contrast to the traditional viewpoint where one would characterize 
the movement of the bands only via their average velocity $\overline{v_b}$. 
Time-averaging each of the fluctuating $v_b(t)$-signals we recover the 
known phenomenology in that $\overline{v_b}$ is found to decrease with the 
strain $\epsilon$ and increase with the strain-rate
$\dot{\epsilon}$ \cite{Klose2004,bands-laserspeckle_shabadi_2004,dic_ait-amokhtar_2006,Jiang2007} (Fig. 2). In this case we found a power-law increase with the strain-rate and an exponential decrease with strain so that they can be summarized 
as
\begin{equation}
    \overline{v_b} \propto \dot{\epsilon}^p \exp \left(- \frac{\epsilon}{\epsilon_0} \right)
\end{equation}
and with the data set at hand, we find $p=0.6$ and $\epsilon_0=0.16$.\\
 
In order to characterize the properties of the $v_b(t)$ signals/velocity
bursts corresponding to individual bands, we start by considering their 
average shapes $\left\langle v_b\left(\frac{t-t_0}{T}\right) \right\rangle$ at a fixed duration/band 
lifetime $T$ (where $t_0$ is the start of the band propagation); this is one of the standard quantities used to characterize
crackling noise bursts. We find that short-lived
bands exhibit an approximately parabolic shape, while considering bands with a 
longer $T$ results in $\left\langle v_b\left(\frac{t-t_0}{T}\right) \right\rangle$ displaying an 
increasingly flattened profile (red symbols in Fig. 
3).\\

How can one theoretically understand the origin and properties of the crackling noise -like $v_b(t)$ band propagation velocity signals, exhibiting such
average temporal velocity profiles? The starting point of
our analysis is the empirical observation that the bands tend to propagate 
essentially as 'rigid bodies', and hence a description based on a single 
degree of freedom, the band position $x_b$, is appropriate. This rigid body 
then moves via overdamped dynamics due to the forces acting on it. 
As the sample is strained with a constant strain-rate 
$\dot{\epsilon}$, it is natural to assume that the 
band position is driven at a rate
$c \propto \dot{\epsilon}$.
This is countered by a stiffness term $k$ which includes the hardening of the sample which can be incorporated in the simplest form as a linear dependence to the strain $k \propto \epsilon$.
As the band propagates along the long axis of the specimen, it samples the 
random dislocation microstructure it encounters during motion, resulting in a 
position-dependent random force $W(x_b)$, with Brownian 
correlations, $\langle W(x_b)W(x_b') \rangle = |x_b - x_b'|$. Collecting these 
terms, one arrives at an equation of motion for $x_b$ which has the same 
form as the Alessandro--Beatrice--Bertotti--Montorsi (ABBM) model \cite{alessandro1990domain} used as the mean-field description of domain 
wall depinning in disordered ferromagnets, i.e.,     
\begin{equation}
\frac{\mathrm{d} x_b}{\mathrm{d} t} = c t - k x_b + D W(x_b) \mathrm{,}
\end{equation}
where $D$ is the disorder strength.
The ABBM model (Eq.~2)
is known to produce crackling noise or avalanches with power-law distributed 
sizes and durations \cite{durin2006science,durin2000scaling}, charaterized by $c$-dependent exponents; for instance,
the size distribution scales as $P(S) \sim S^{-(3-\tilde{c})/2}$ where $\tilde{c} = c/D$ is the normalized driving rate. 
Following Ref. \cite{papanikolaou2011universality}, Eq.~2 can be transformed to a form including 
a time-dependent noise term,
\begin{equation}
\frac{\mathrm{d} v_b}{\mathrm{d} t} = c - k v_b +\sqrt{2 D v_b} \xi(t),
\end{equation}
with $\xi$ being a white noise term with unit variance $\langle \xi(t) \xi(t') \rangle = \delta(t-t')$. This Eq.~3 has
the advantage of allowing one to solve analytically quantities like the 
average burst shape in the $k=0$, $c=0$ limit, resulting in an inverted parabola, 
while a finite $k$ gives rise to a flattening of the shape for long 
avalanches \cite{papanikolaou2011universality}.\\

\paragraph*{Comparing the model and the experiments}
To compare this model with experimental data, we simulate 
it by ``nucleating bands'' at random
initial positions $x_b^i \in (0,L)$ within a sample of length $L$, and let them
propagate in a random direction according to Eq.~3. To mimic effects due to the finite length of the sample, we consider only bands 
that stop before the end of the sample. This leads to a 
$L$-dependent cutoff to the ``avalanche'' distribution; in addition, a cutoff could in principle be due to $k$ in Eq.~3, but here
$k$ is sufficiently small such that the $L$-dependent cutoff dominates. This then results in a scaling form for the avalanche size distribution
(see Methods for details)
\begin{equation}
P(S) \propto \left(1-\frac{S}{L}\right) S^{-(3-\tilde{c})/2} \mathrm{.}
\end{equation}\\

Fixing the model parameters to 
$c=c_0 \dot{\epsilon}$ (where $c_0 = 50$ m/s), $k=k_0 \epsilon$ (where $k_0 = 650 \; \mathrm{s}^{-1}$) and $D=700 \; \mathrm{mm}/\mathrm{s}^3$
reproduces the experimentally seen band velocities well (Fig. 2), both in terms of the behavior with increasing strain-rate and strain. 
The number of bands generated with a given strain and strain-rate is proportional to the statistics of the experiments.
All this is achieved with a very simple linear proportionality of the model parameters to the experimental parameters $\epsilon$ and $\dot{\epsilon}$.\\

The model also leads to an excellent agreement in terms of the velocity profiles (Fig. 3): inverted parabola for small $T$ and increasingly flattened shape for larger $T$. The slight asymmetry \cite{zapperi2005signature,laurson2013evolution} seen in the experimental profiles, where the velocities are slightly larger towards the start of the band lifetime, is naturally not reproduced by the model.\\

We then proceed to a more extensive statistical characterization of the
band propagation velocity bursts, again exploiting the remarkable analogy
with standard crackling noise systems.
First we look at the scaling of the avalanche size $S$ with the duration $T$ in the experiments and in the simulations (where the parameters were fitted to reproduce the band velocity behavior with increasing strain and strain-rate).
Typically one would expect $\langle S \rangle \sim T^2$  \cite{durin2006science,durin2000scaling} from the ABBM model but here the finite sample of length $L=28$ mm (and the exclusion of bands with $S < 2$ mm, see Methods for details) restricts this scaling. Additionally with different strains and strain-rates one would have different prefactors and averaging over these would not result in the typical scaling form. Indeed after performing this average we see very similar behavior of $\langle S \rangle$ increasing with duration (Fig. 4a) for both the experiments and the ABBM model and the increase is slower than $\langle S \rangle \sim T^2$.\\

To connect the predictions of the ABBM model to the stress-strain curve one can also study the the scaling of the average stress-drop size $\langle \Delta \sigma \rangle$ from the stress-strain curves with duration. The one-to-one correspondence between deformation bands and stress-drops is broken by the observed multiple simultaneously propagating bands (Fig. 1)
but the average scaling seems to be similar for the avalanche sizes and stress-drop sizes (Fig. 4a). The shorter stress-drop durations are due to the simultaneous bands and the short $S < 2$ mm bands that are otherwise neglected from the analysis.\\

The prediction for the avalanche size distribution in a finite-size sample (Eq.~4) has a driving rate dependent exponent $(3-\tilde{c})/2$. However as we are again observing bands at different strains and strain-rates it is easier to consider a distribution $P(S) \propto \left( 1-\frac{S}{L} \right) S^{-\alpha}$ with some exponent $\alpha$. The experimental data seems to follow this distribution quite well (Fig. 4b) and maximum likelihood estimation gives $\alpha = 0.99$.
The same is true for the ABBM model where a slightly lower estimate of $\alpha = 0.73$ is obtained. This disparity in the exponent values is likely due to the simplicity of the model parameters (linearity of $c$ and $k$ in $\dot{\epsilon}$ and $\epsilon$) and the fact that the parameter values were fitted just to reproduce the behavior with increasing strain and strain-rate.\\

The instantaneous (band) velocity distribution in the ABBM model is known to be of the form \cite{alessandro1990domain,zapperi1998dynamics,papanikolaou2011universality,leblanc2012distribution}
\begin{equation}
    P(v_b) = \frac{\tilde{k}}{\Gamma(\tilde{c})} v_b^{\tilde{c}-1} \exp\left( - \tilde{k} v_b \right)
\end{equation}
where $\tilde{k} = k/D$ is the normalized stiffness term and $\Gamma$ represents the Gamma function. As the observed size distribution suggests $\tilde{c}$ to be around unity one would expect an exponential band velocity distribution $P(v_b) = \tilde{k} \exp\left( - \tilde{k} v_b \right)$. Indeed this is what we see in Fig.~4c where the distribution of experimental band velocities follows an exponential distribution for velocities between 20 mm/s and 110 mm/s. For the velocities obtained from the ABBM model the exponential distribution is observed for velocities from 20 mm/s to 200 mm/s.
The parameter $\tilde{k}$ depends on strain but the best fit to the distributions is obtained with $\tilde{k} \approx 0.04$.

\section*{Discussion}

We have studied the statistics of the PLC deformation bands by using a fast imaging technique and by a simple mean-field model of avalanche dynamics. Empirically, we found that the average band velocities scale on average as
$\overline{v_b} \propto \dot{\epsilon}^p \exp\left( - \frac{\epsilon}{\epsilon_0} \right)$ with $p=0.6$ and $\epsilon_0 = 0.16$
and exhibit a flattening of the average velocity profile $\left\langle v_b\left(\frac{t-t_0}{T}\right) \right\rangle$ with increasing band duration.
Remarkably these features can be reproduced with the ABBM model by taking the strain-rate to represent the driving term and strain to represent the stiffness term, the strain hardening. The material properties are also contained in the disorder strength. Another key idea is constraining the simulated bands to the finite size of the sample in order to match the statistics with the experiments.\\

The finite size of the sample and the dependence of the model parameters on both strain and strain-rate hide the known scaling form of the avalanche size. However our simulations show that both the average PLC band propagation distances and ABBM avalanche sizes scale similarly with duration. The same scaling can also be seen for the stress-drop sizes in the stress-strain curve. Although the one-to-one correspondence between bands and stress-drops is lost with multiple simultaneous bands, the average scaling remains the same.\\

We analytically show that the finite size of the sample introduces a $\left(1-\frac{S}{L}\right)$-cutoff to the known power-law avalanche size distribution. Both the PLC band propagation distances and the simulated bands from the ABBM model follow this $P(S) \sim \left( 1 - \frac{S}{L} \right) S^{-\alpha}$ distribution with exponents $\alpha$ close to unity. Based on this one would then expect the instantaneous band velocity distribution to follow an exponential distribution, which is what we see in the experiments and in the simulations.
The ABBM model is commonly studied close to the quasistatic limit $c \to 0$, however we have showed here that it can be also used to explain the behavior of fronts under strong drive, here the deformation bands. Looking at these
results from the viewpoint of classical theories of the PLC, it is an important question how to modify and adapt such models of DSA \cite{Hahner2001} so that they reproduce correctly the kind of stochasticity seen in the band dynamics. This may be restated so that the "correct" model one should be able to reduce to the ABBM used here.  \\

What our results show is that interacting, mobile dislocations create avalanches of deformation in metal alloys. Here, the
necessary conditions for this are 
temperature and strain-rate values within a specific window
such that propagating, or type A PLC bands are observed.
Given this, the avalanches follow the paradigm of the mean-field-like ABBM model. The eventual stopping of the band is a random fluctuation, and depends on the local, heterogeneous material properties. Thus the physics of these bands arises from a mixture of external drive, local randomness, and the coarse-grained, collective response of many dislocations. More work is needed in understanding the implications to other PLC band types, and what the practical predictions or consequences are for alloys with different composition ("disorder") and for samples of different sizes. It is likely that the ABBM exponent $\tilde{c}$ is material-dependent.
A wider look suggests to consider the eventual interaction physics of multiple bands present in the sample, where their interaction with others and with the sample or disorder would be crucial \cite{reichhardt2016depinning}.
In the same vein, propagating bands of deformation with serrations of the stress-strain curve are also seen in the plastic deformation of amorphous materials \cite{ramachandramoorthy2019dynamic,li2020temperature}. An obvious question would be if these also can be shown to follow ABBM-like dynamics with a careful study, but then again if such bands do not follow this simplest paradigm that is also of profound interest.

\section*{Materials and Methods}

\paragraph*{Experimental methods}
The laser speckle technique \cite{bands-laserspeckle_shabadi_2004} 
was used to observe the bands 
in a commercial aluminum alloy AW-5754 sample. The samples were lasercut to a flat dogbone shape 
with the dimensions 28 mm $\times$ 4 mm $\times$ 0.5 mm for the gauge volume.
The samples have a polycrystalline structure with an average grain size of 38 $\pm$ 14 $\mu$m. The experimental setup 
is illustrated in Fig. 5.\\

The samples were tensile loaded with Instron ElectroPuls E1000 using an Instron Dynacell load cell with a constant displacement rate. The stress and strain were calculated from the displacement and force data provided by the machine. These were recorded with an acquisition rate of 500 Hz and the samples were held using an initial force of 4 N.\\

The speckle pattern was recorded with ProtoRhino FlexRHINO DynaMat system which includes a high speed camera, 
a laser and a FPGA-chip based unit for data acquisition and storage. The camera had an electronic freeze-frame 
shutter and a Navitar MVL7000 objective with a macro zoom lens, an aperture of f/2.5 and a spatial resolution 
of 54 $\mu$m. The laser used was a collimated laser diode with a wavelength of 638 nm and a power of 200 mW.
The acquisition rates varied around 0.5-2.0 kHz.\\

The speckle images were analyzed using the equal interval subtracting method (similar to Ref. \cite{bands-laserspeckle_shabadi_2004}) where the subtraction was done for consecutive images or with the highest acquisition rate for every other image. A 1D projection was taken from these subtracted images in direction perpendicular to the band with the two different band inclinations. This provides two different effective strain-rate maps where the measured quantity $\dot{\epsilon}_{spec}$ corresponds to the time derivative of the speckle image intensity.\\

As the band angles and widths were observed to remain very close to constant (the band widths are 0.9 $\pm$ 0.1 mm which is of the order of the sample thickness 0.5 mm) these effective strain-rate maps were used to track the band movement as a 1D rigid body. The maximum value of $\dot{\epsilon}_{spec}$ around the visible band corresponds to the leading edge of the band and this was used as the band position $x_b$ therefore also determining the propagation distance $S$ and band duration $T$. As we are considering type A band dynamics a propagation distance cutoff of 2 mm was imposed. Bands that propagate less than 2 mm correspond more to the type B regime of nonpropagating bands and were excluded from the analysis. The band velocity signal $v_b$ was then obtained by numerically differentiating the band position signal and the average band velocity was calculated simply as $\overline{v_b} = S/T$.\\

There are sometimes multiple simultaneous bands present in the sample (see Fig.~1a and Fig.~1c) which can collide with each other. These collisions usually lead to the disappearance of both bands (except for a few cases right before the sample failure where bands with different inclinations can interact, for more details see Supplementary video) and therefore we have chosen to consider each band as an independent event.

\paragraph*{Fitting the model parameters}
The model parameters $c$, $k$ and $D$ were fitted to the experimental data in the following way. First a value for $c$ was chosen high enough (with some arbitrary $k$ and $D$) so that the average band velocities $\langle \overline{v_b} \rangle$ behave with the correct exponent (or slope) as a function of the strain-rate $\dot{\epsilon}$. After that the value of $k$ was chosen so that the normalized average band velocities $\overline{v_b} \dot{\epsilon}^{-p}$ behave with the correct slope as a function of the strain $\epsilon$. If the behavior of $\overline{v_b} \dot{\epsilon}^{-p}$ had a strong strain-rate dependence $c$ was increased and the value for $k$ chosen again. Finally the value for $D$ was chosen so that that the actual values (not just the slopes) of $\overline{v_b} \dot{\epsilon}^{-p}$ best fit the experimental data.

In the end all three of the values were perturbed around the chosen values to make sure that the values represent a local minima of the difference between the simulation results and experimental data.\\

\paragraph*{Derivation of the avalanche size distribution}
One can calculate the effect of the finite sample length on the avalanche size distribution or in other words the band propagation distance by a simple construction. Nucleating bands at a random position on the one-dimensional sample, letting them propagate and constraining the propagation to the size of the sample gives directly the cutoff induced by the finite size sample.\\

Let $Y$ be the event of a band starting at a certain position $x_b^i \in (0, L)$ and $X$ be the event of the band stopping at a certain position $x_b^f \in (0, L)$. As we see in the experiments a flat distribution of the starting positions the joint probability is then
\begin{equation}
    P_{XY}(x_b^f, x_b^i) = P_{X | Y}(x_b^f, x_b^i) P_Y(x_b^i) = \frac{P_{X | Y}(x_b^f, x_b^i)}{L}
\end{equation}
and one can get the distribution of the travel distance $S$ by calculating two convolutions
\begin{equation}
    P_S(s) =
    \int_{x_b^f=0}^{x_b^f=L-s} P_{X | Y}(x_b^f, x_b^f+s) P_Y(x_b^f+s) \, \mathrm{d} x_b^f
    + \int_{x_b^f=s}^{x_b^f=L} P_{X | Y}(x_b^f, x_b^f-s) P_Y(x_b^f-s) \, \mathrm{d} x_b^f \mathrm{.}
\end{equation}\\

The conditional probability is handled most simply by splitting it into two portions.
After starting the band goes in either direction with equal probability and as is known for the ABBM model travels a distance that is power-law distributed.
The joint distribution is then
\begin{equation}
    P_{X|Y}(x_b^f, x_b^i) \propto 
    \begin{cases} (x_b^i-x_b^f)^{-\alpha}, & x_b^f < x_b^i \\ (x_b^f-x_b^i)^{-\alpha}, & \mathrm{otherwise}
    \end{cases}
\end{equation}
and the convolutions give
\begin{equation}
    P_S(s) \propto 
    \int_{x_b^f=0}^{x_b^f=L-s} s^{-\alpha} \, \mathrm{d} x_b^f
    + \int_{x_b^f=s}^{x_b^f=L} s^{-\alpha} \, \mathrm{d} x_b^f
    \propto \left(1-\frac{s}{L}\right) s^{-\alpha} \mathrm{.}
\end{equation}
Normalizing this distribution (from a minimum value $S_0$ to $L$) gives the full functional form
\begin{equation}
    P(S) = \frac{\left(1-\frac{S}{L}\right) S^{-\alpha}}{\frac{L^{1-\alpha}-S_0^{1-\alpha}}{1-\alpha}
    - \frac{1}{L} \frac{L^{2-\alpha}-S_0^{2-\alpha}}{2-\alpha}}
\end{equation}
or in the special case of $\alpha = 1$
\begin{equation} \label{eq:sizeAlphaOne}
    P(S) = 
    \frac{\left(1-\frac{S}{L}\right) S^{-1}}{\frac{S_0}{L} + \ln \frac{L}{S_0} - 1} \mathrm{.}
\end{equation}

For the special case of $k=0$ one can obtain an analytic solution for the exponent in the ABBM model as $\alpha = \frac{3-\tilde{c}}{2}$ \cite{durin2006science,durin2000scaling} 
which would here give the avalanche size distribution
\begin{equation}
    P(S) = \frac{1}{2} \frac{\left(1-\frac{S}{L}\right) S^{-(3-\tilde{c})/2}}{\frac{L^{(\tilde{c}-1)/2}-S_0^{(\tilde{c}-1)/2}}{\tilde{c}-1}
    - \frac{1}{L} \frac{L^{(\tilde{c}+1)/2}-S_0^{(\tilde{c}+1)/2}}{\tilde{c}+1}}
\end{equation}
or in the case of $\tilde{c} = 1$ the one shown in Eq.~11.





\section*{Acknowledgments}
We thank Ivan Lomakin for the determination of the sample grain structure.
\paragraph*{Funding:} We acknowledge the financial support from the
Academy of Finland through the Centers of Excellence
program (Project No. 251748), an Academy Research
Fellowship (L.L., Project No. 268302), and the Academy Projects COPLAST (L.L., Project No. 322405) and FLUFRA (M.A., Project No. 317464). T.M. acknowledges the support of The Finnish Foundation for Technology Promotion. 
M.A. acknowledges support from the European Union Horizon 2020 research and innovation
programme under grant agreement No. 857470 and from European Regional Development Fund via Foundation for Polish Science International Research Agenda PLUS programme grant No. MAB PLUS/2018/8.
We acknowledge the computational resources provided by the Aalto 
University School of Science “Science-IT” project.
\paragraph*{Competing interests:}
The authors declare that they have no competing interests.
\paragraph*{Author contributions:}
T.M., P.K. and M.O. performed the experiments and T.M. analyzed the data and performed the simulations. T.M., L.L., and M.J.A. wrote the manuscript.
\paragraph*{Data availability:}
All data needed to evaluate the conclusions in the paper are present in the paper. The data that support the findings of this study are 
available from the corresponding authors on reasonable request.

\section*{Supplementary materials}
Supplementary Video 1: An example video of speckle images from an experiment showing the complex dynamics of the deformation bands from the yielding of the sample until the final failure. The color corresponds to the speckle image intensity (from dark to light) which corresponds to the local strain rate.\\


\clearpage

\begin{figure}[!]
\centering
\includegraphics[width=0.48\textwidth]{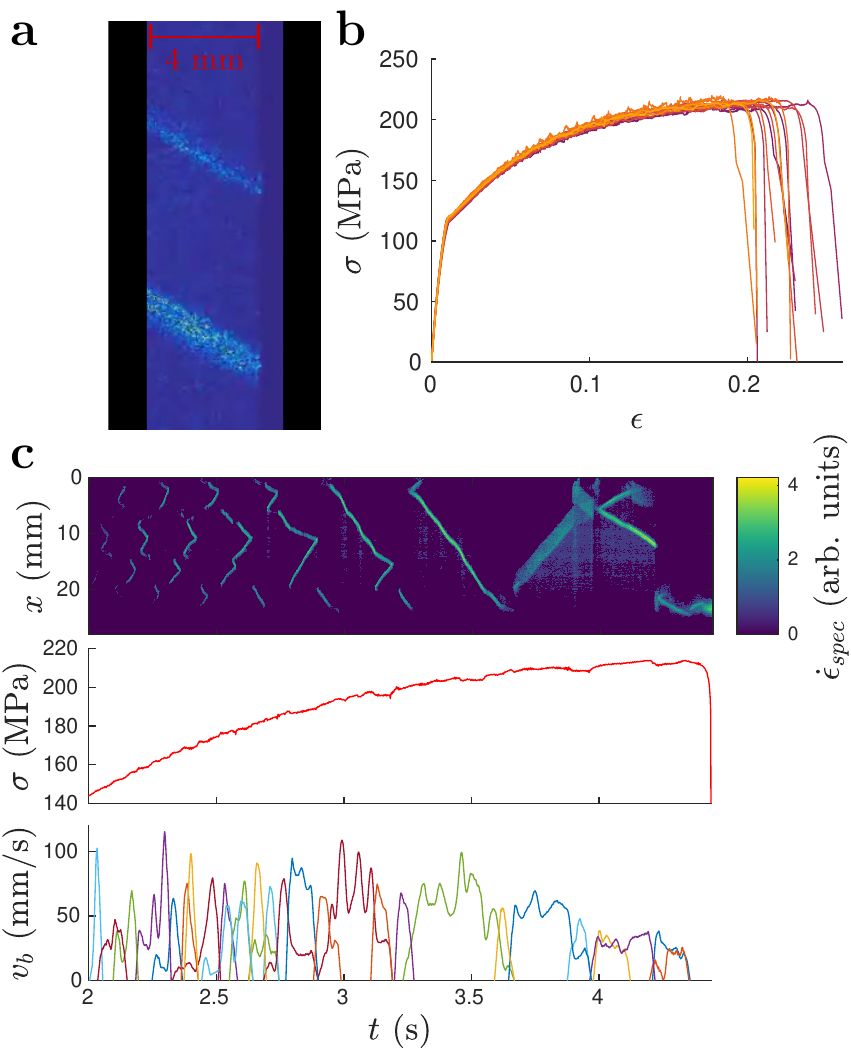}
\caption{{\bf Features of deformation bands from speckle images.}
({\bf a}) The raw subtracted speckle images showing two simultaneous PLC bands (top one during nucleation). 
({\bf b}) The stress-strain curves showing the responses of the samples and the serrations. 
({\bf c}) The effective strain-rate map (time derivative of the speckle image intensity) $\dot{\epsilon}_{spec}$ (see Methods for details) for one band inclination (top),
the stress signal (middle) and the band velocity signals (bottom).}
\end{figure}

\begin{figure}[!]
\centering
\includegraphics[width=0.48\textwidth]{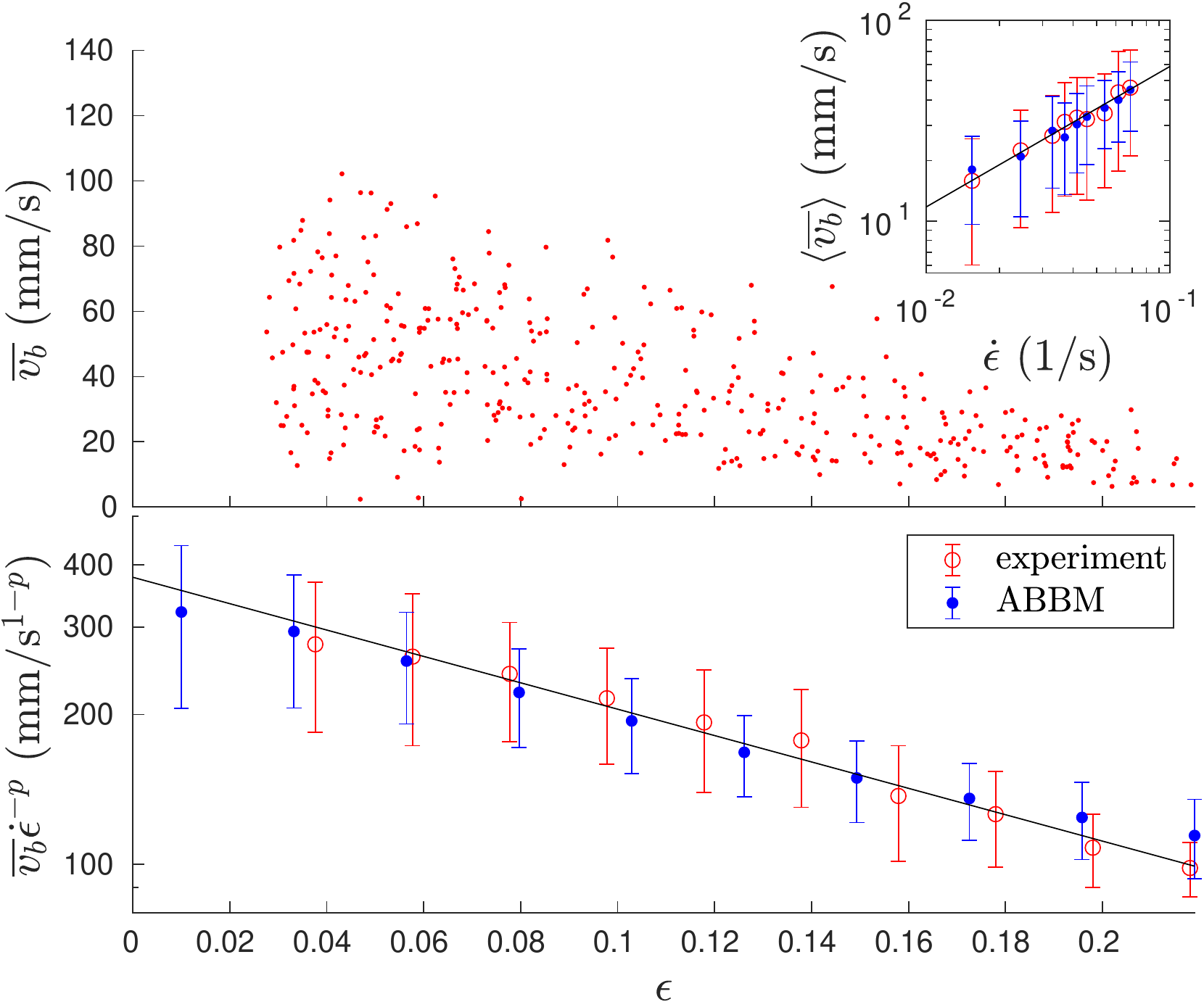}
\caption{{\bf Average band velocities.}
\emph{Top:} The average velocities $\overline{v_b}$ of the bands decrease with strain $\epsilon$. The Inset shows the average velocities averaged over the strain-rate $\left\langle \overline{v_b} \right\rangle$ increasing as a power-law (red symbols for experiments and blue for the ABBM model, the error bars representing the standard deviation of band velocities obtained with a given strain-rate). The black line is a power-law $\left\langle \overline{v_b} \right\rangle \sim \dot{\epsilon}^p$ with $p=0.6$.
\emph{Bottom:} The average velocities (binned to strain intervals) scaled with $\dot{\epsilon}^p$ decrease exponentially with strain (red symbols for experiments and blue for the ABBM model, the error bars representing the standard deviation of band velocities in the given strain bin).
The black line is an exponential relation $\overline{v_b} \dot{\epsilon}^{-p} \sim \exp\left( -\frac{\epsilon}{\epsilon_0} \right)$ with $\epsilon_0=0.16$.}
\end{figure}

\begin{figure}[!]
\centering
\includegraphics[width=\textwidth]{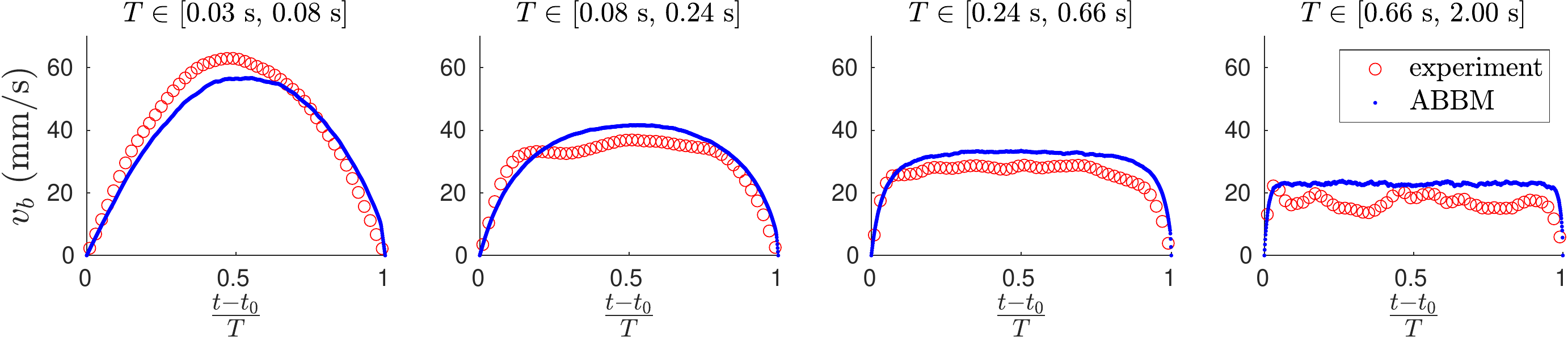}
\caption{{\bf Averaged band velocity profiles for a fixed duration.}
The averaged band velocity profiles for four different duration bins (red) showing the evolution of the shape from an inverted parabola to almost a flat constant velocity shape with increasing duration. As expected from the analytic results this is reproduced by the ABBM model (blue).}
\end{figure}

\begin{figure}[!]
\centering
\includegraphics[width=0.48\textwidth]{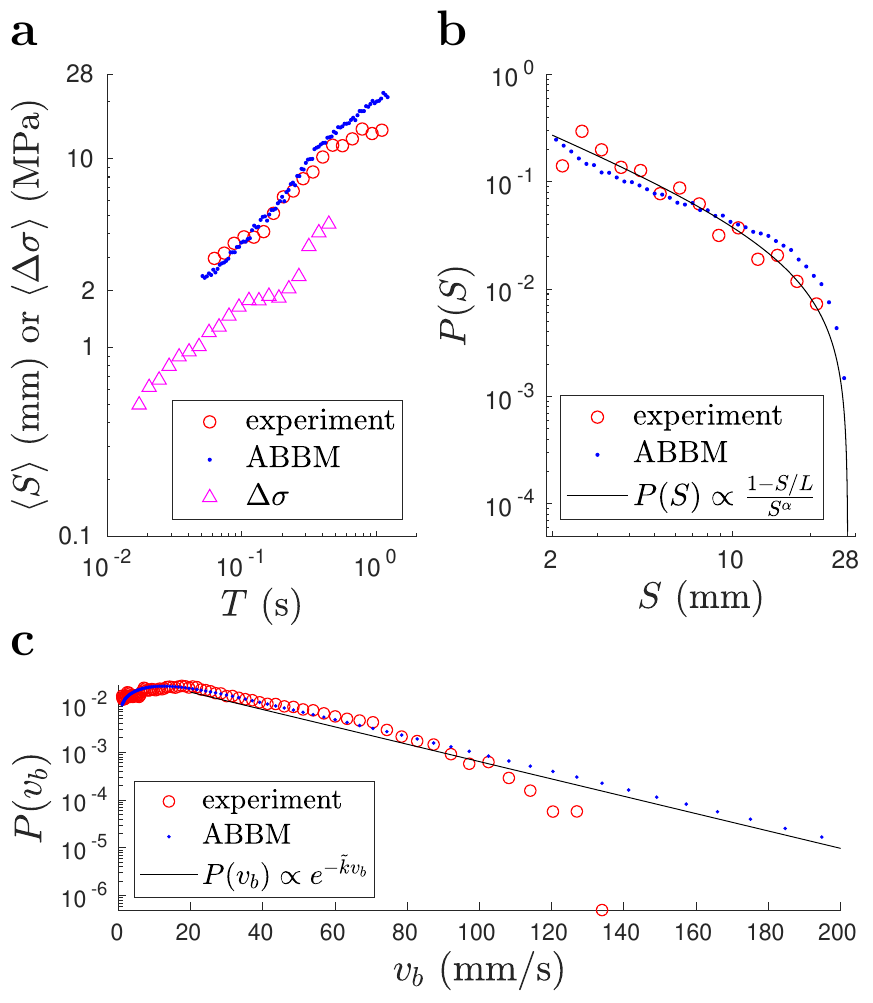}
\caption{{\bf Statistics of the avalanche sizes.}
({\bf a}) The average avalanche size $\left\langle S \right\rangle$ for a fixed duration $T$ for the experiments (red) and the ABBM model (blue). The third curve (magenta) is the fixed duration average for the size of the stress-drop $\Delta \sigma$ in the serrated stress-strain curve. ({\bf b}) The avalanche size distribution from the experiments (red) and the ABBM simulations (blue). The black line represents the expected scaling of the distribution $P(S) = A \left(1 - \frac{S}{L} \right) S^{-\alpha}$ with $\alpha=1$, $A = \left( \frac{S_0}{L} + \ln \frac{L}{S_0} - 1 \right)^{-1}$ and $S_0 = 2$ mm. ({\bf c}) The distribution of band velocities from the experiments (red) and the ABBM simulations (blue). The black line represents the expected exponential distribution $P(v_b) = \tilde{k} e^{-\tilde{k} v_b}$ with $\tilde{k}=0.04$.}
\end{figure}

\begin{figure}[!]
\centering
\includegraphics[width=0.48\textwidth]{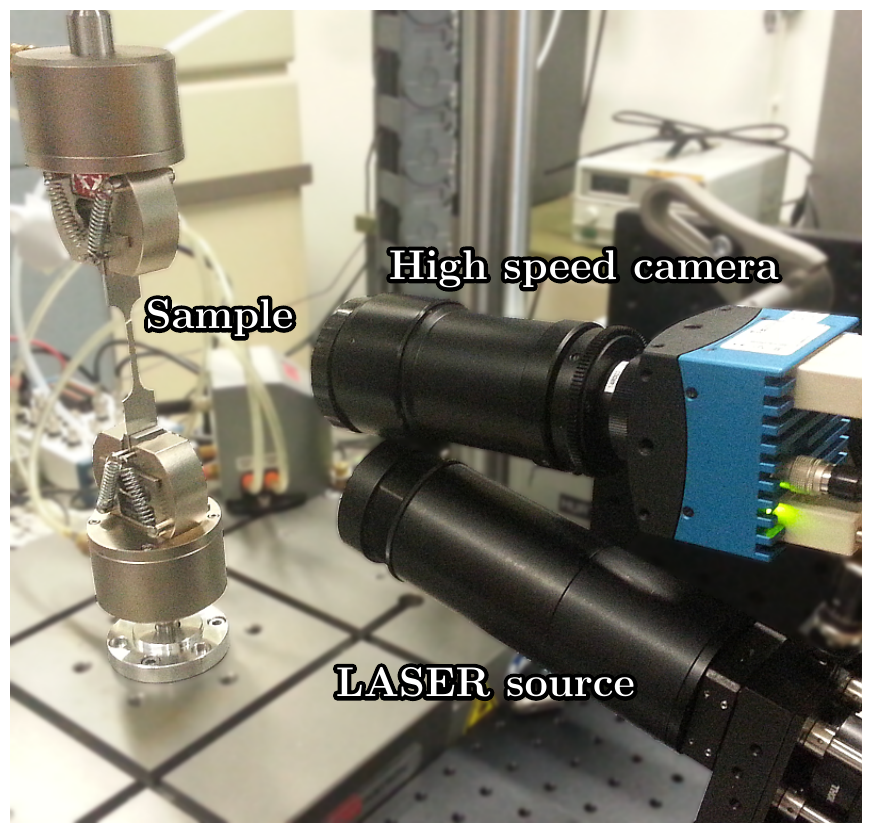}
\caption{{\bf The experimental setup.}
The sample is tensile loaded and simultaneously imaged using the laser speckle technique. Here the sample is illuminated by a diffuse laser at a slight angle and the produced speckle pattern on the sample surface is imaged using a high speed camera. Photo credit: Tero Mäkinen}
\end{figure}

\end{document}